\documentclass[aps,prb,superscriptaddress,twocolumn]{revtex4}
\usepackage[pdftex]{graphicx}

\begin{document}

\title{Coupling and coherent electrical control of two dopants in a silicon nanowire }
\author{E. Dupont-Ferrier}
\affiliation{SPSMS, UMR-E CEA / UJF-Grenoble 1, INAC, Grenoble, F-38054, France}
\author{B. Roche}
\affiliation{SPSMS, UMR-E CEA / UJF-Grenoble 1, INAC, Grenoble, F-38054, France}
\author{B. Voisin}
\affiliation{SPSMS, UMR-E CEA / UJF-Grenoble 1, INAC, Grenoble, F-38054, France}
\author{X. Jehl}
\affiliation{SPSMS, UMR-E CEA / UJF-Grenoble 1, INAC, Grenoble, F-38054, France}
\author{R. Wacquez}
\affiliation{CEA/LETI-MINATEC, CEA-Grenoble}
\author{M. Vinet}
\affiliation{CEA/LETI-MINATEC, CEA-Grenoble}
\author{M. Sanquer}
\affiliation{SPSMS, UMR-E CEA / UJF-Grenoble 1, INAC, Grenoble, F-38054, France}
\author{S. De Franceschi}
\affiliation{SPSMS, UMR-E CEA / UJF-Grenoble 1, INAC, Grenoble, F-38054, France}

\maketitle

{\bf

Electric control of individual atoms \cite{Kane1998,Vrijen2000,Hollenberg2004} or molecules\cite{Bogani2008} in a solid-state system offers a promising way to bring quantum mechanical functionalities into electronics. This idea has recently come into the reach of the established domain of silicon technology, leading to the realization of single-atom transistors\cite{Sellier2006,Ono2007,PierreM.2010,Fuechsle2012} and to the first measurements of electron spin dynamics in single donors\cite{Morello2010}. Here we show that we can electrically couple two donors embedded in a multi-gate silicon transistor, and induce coherent oscillations in their charge states by means of microwave signals. We measure single-electron tunneling across the two donors, which reveals their energy spectrum. The lowest energy states, corresponding to a single electron located on either of the two donors, form a two-level system  (TLS) well separated from all other electronic levels. Gigahertz driving of this TLS results in a quantum interference pattern associated with the absorption or the stimulated emission of up to ten microwave photons. We estimate a charge dephasing time of 0.3 nanoseconds, consistent with other types of charge quantum bits\cite{Nakamura2002,Hayashi2003,Petta2004}. Here, however, the relatively short coherence time can be counterbalanced by fast operation signals (in principle up to 1 terahertz) as allowed by the large empty energy window separating ground and excited states in donor atoms. The demonstrated coherent coupling of two donors constitutes an essential step towards donor-based quantum computing devices in silicon.

}

Donor atoms in silicon (e.g. P or As) draw increasing attention in view of their use as qubits for the development of quantum computing devices based on silicon technology\cite{Kane1998,Vrijen2000,Hollenberg2004,Morton2011}. Quantum information can be encoded either in the charge state of an electron confined in the double-well potential of two nearby donors\cite{Hollenberg2004}, or it can be encoded in the 
electron \cite{Vrijen2000} or the nuclear spin state \cite{Kane1998} of a single donor. 
While charge qubits allow for simpler and faster schemes for qubit manipulation and read-out, spin qubits offer longer quantum coherence\cite{Tyryshkin2012}. Realizing and controlling the tunnel coupling between two donor atoms is a basic requirement not only to create a double-donor charge qubit, but also to control entanglement of adjacent single-donor spin qubits\cite{Kane1998,Hollenberg2004}. In the latter case, entanglement is provided by a tunnel-mediated exchange interaction between the spins of two donor electrons. Here we demonstrate coherent electrical control of double-donor systems in silicon and use Landau-Zener-Stuckelberg interferometry\cite{Shevchenko2010} to probe the coherence of charge oscillations between two donors.

\begin{figure}
\begin{center}
\includegraphics[width=0.8\columnwidth]{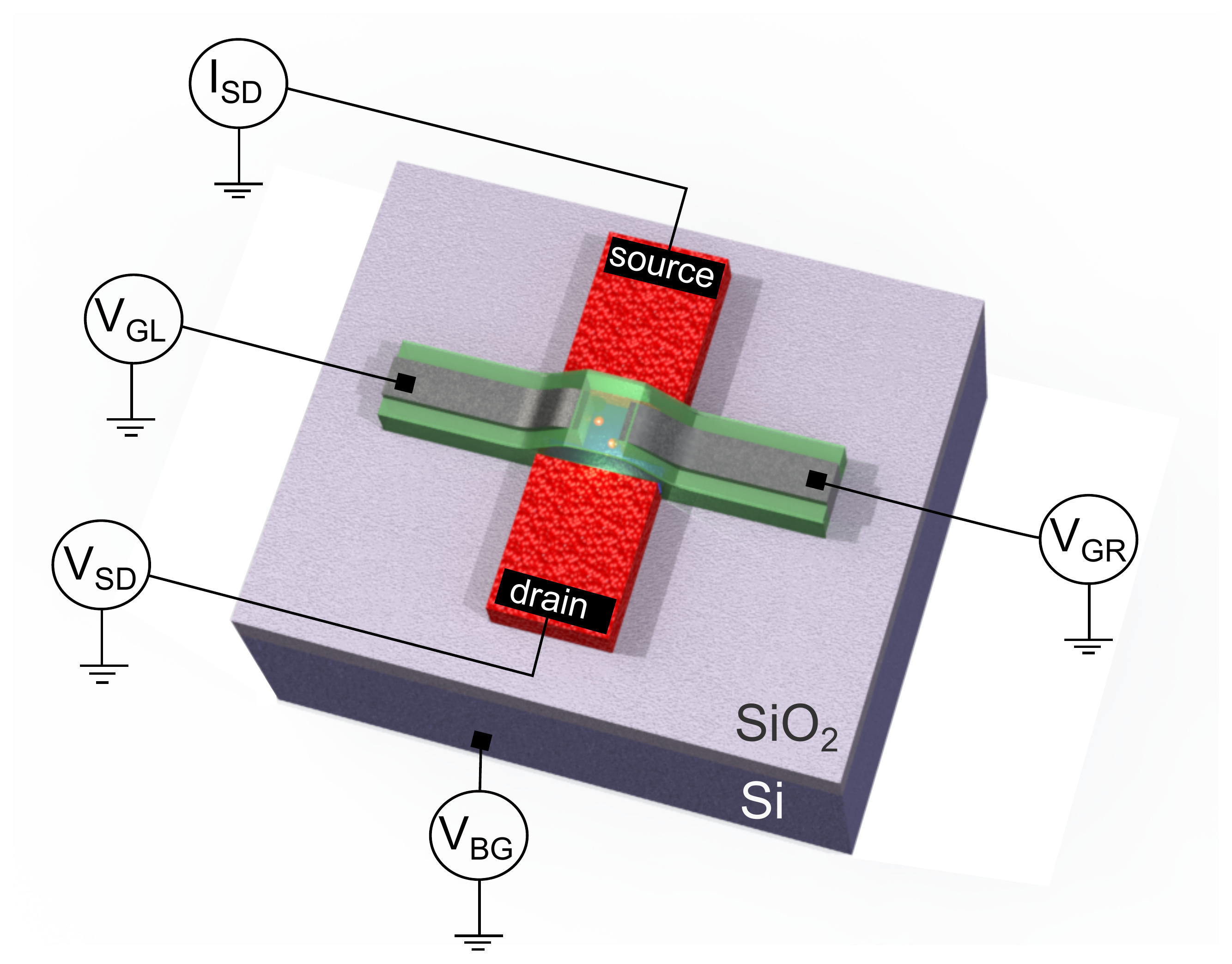}
\caption{\textbf{Device layout, electrical wiring and measurement scheme}. 
The device consists of a 20-nm-thick, 60-nm-wide silicon nanowire etched out of a silicon-on-insulator substrate and partially overlapped in its center by two facing polysilicon top gates (in gray). The silicon substrate acts as an additional back gate. The silicon nanowire is implanted with As donors (red spheres) creating degenerately doped source and drain leads, while the gate electrodes and the surrounding insulating spacer layers (green) protect the 40\,nm long channel region from high-dose implantation. Different methods are used to implant a few donor atoms (As or P) in the channel region (see Supplementary Material). Transport through these few dopants is accomplished by independently tuning their levels into the energy window set by the source-drain bias ($V_{SD}$). This tuning is accomplished by means of the two top gate voltages ($V_{GL}$, $V_{GR}$) and a back-gate voltage ($V_{BG}$). A DC current $I_{SD}$ through the dopants is measured. For experiments presented in Fig. 3 and Fig. 4 an additional microwave signal is applied to one of the top gates.} 
\end{center}
\end{figure}

Experiments were carried out on silicon nanowire transistors provided with multiple gates as shown in Fig.1. Doping impurities were introduced in the silicon channel by conventional ion implantation techniques. This process was tuned to yield a weakly doped transistor channel, containing just a few dopants, sandwiched between degenerately doped (i.e. metal-like) silicon leads\cite{Pierre2009c} (see Supplementary Material). 
We performed low-temperature (15 mK) transport measurements in different samples implanted with either As or P donors, as well as in control samples with no intentional channel doping. Only in channel-implanted devices could tunnel transport through channel donors be achieved. The multi-gate device geometry allowed the selection of the conductive path corresponding to tunneling through pairs of donor atoms connected in series\cite{Larkin1987}. This transport regime was achieved by tuning the transistor into its pinch-off state, close to the onset of free-carrier conduction (see Supplementary Material).

Fig. 2a shows the source-drain current, $I_{SD}$, of an As-implanted device measured as a function of the two top gates voltages, $V_{GL}$ and $V_{GR}$, for a source-drain bias $V_{SD}=15$ mV.
Current flow occurs inside two overlapping triangular regions with the same size and orientation, a clear signature \cite{Wiel2003} of transport through two dopants in series.
To facilitate the analysis of the observed current features we label the donor atom closer to gate $G_L$ (resp. $G_R$) as $D_L$ (resp. $D_R$) . We denote the corresponding charge states ($N_L$,$N_R$) where $N_L$ ($N_R$) is the electron occupancy of $D_L$ ($D_R$). In this notation, current flow within the lower triangle of Fig. 2a is accomplished through the ($N_L$,$N_R$) charge cycle: (0,0) $\rightarrow$ (1,0)  $\rightarrow$  (0,1)  $\rightarrow$ (0,0). Instead, the upper triangle is associated with the ($N_L$,$N_R$) charge cycle: (1,1)  $\rightarrow$ (1,0)  $\rightarrow$  (0,1)  $\rightarrow$ (1,1). Each of the above cycles results in the transfer of one electron from source to drain \cite{Wiel2003}. The offset between the corresponding triangles reflects the energy cost of the Coulomb repulsion between the donor electrons. We measure an electrostatic energy difference of 4.5 meV between the (1,1) and the (0,0) state corresponding to a dopant separation \cite{Shklovskii1984} of 30 nm.

\begin{figure}
\begin{center}
\includegraphics[width=0.8\columnwidth]{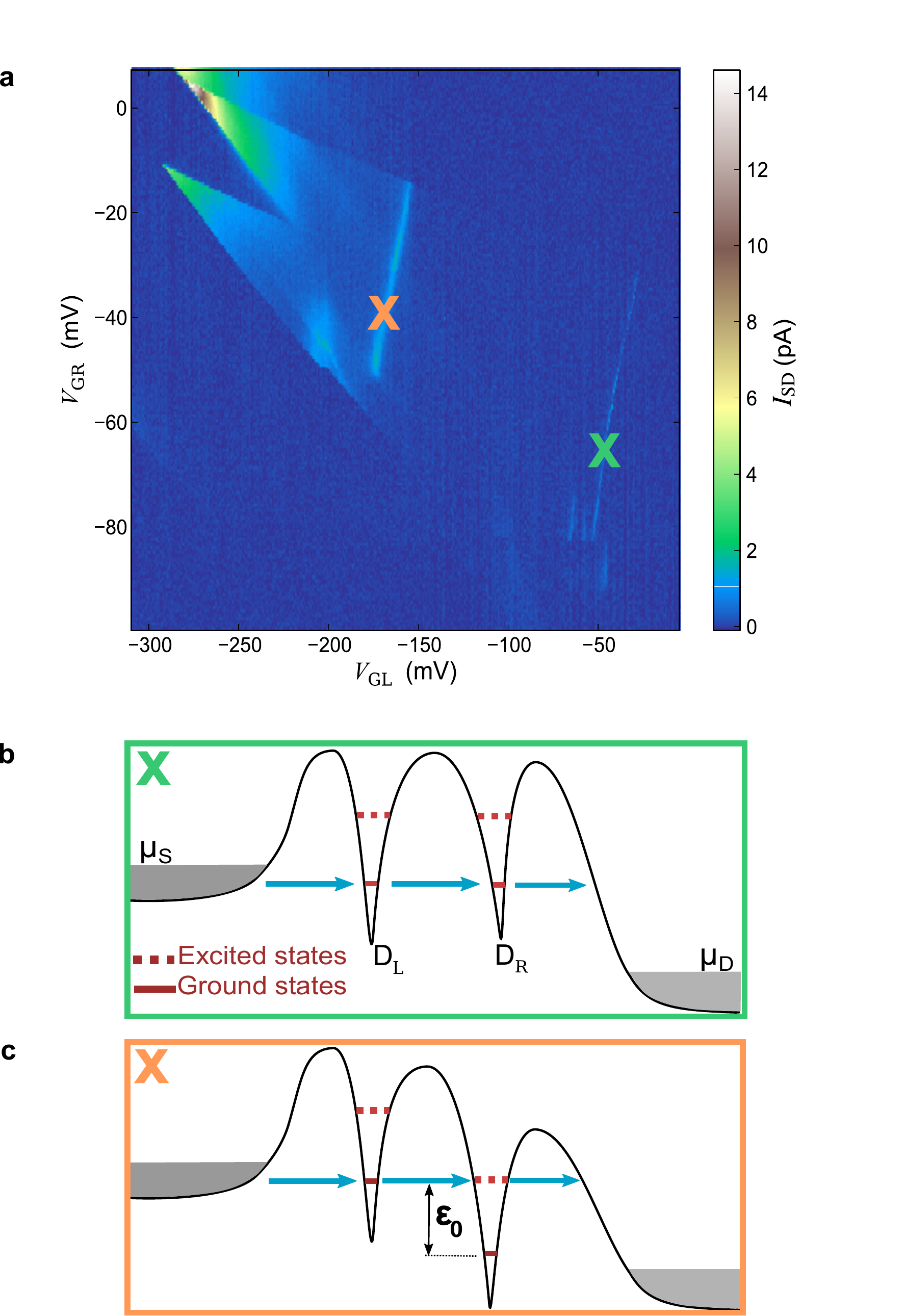}
\caption{\textbf{DC transport measurement through two As donors connected in series.}
\textbf{a} Current map, $I_{SD}$($V_{GL}$,$V_{GR}$), for a fixed $V_{SD}= 15 mV$. At 15 mK, the observed current features, are confined within two triangular-shaped regions, denoting the sequential tunneling of single electrons through a pair of As donors in series, labeled as $D_L$ and $D_R$. $V_{GL}$ and $V_{GR}$ allow for a full control of the dopant energy levels. $\epsilon_0$ defines the level detuning between the donor ground-states. $\epsilon_0 = 0$ all along the bases of the triangles (green cross) where a current peak is observed due to resonant tunneling via the donor ground states. $\epsilon_0 = eV_{SD}$ at the apex of each triangle, providing the scaling factor between gate voltage and energy. 
A second current ridge is observed parallel to the base line (orange cross). This resonance is due to the alignement of $D_L$'s ground state with $D_R$'s first excited state. Its distance from the base line corresponds to $\epsilon_0 = 7.4 \pm 0.4$ meV yielding a direct measurement of the excitation energy for donor $D_R$. We note that no measurable current is detected between ground- and excited-state lines, indicating a minor contribution of inelastic tunneling between the donor ground states. \textbf{b} and \textbf{c} illustrate the qualitative energy diagrams corresponding to the green and orange crosses in \textbf{a}.
} 
\end{center}
\end{figure}

The energy positions of the donor ground-state levels are independently controlled by $V_{GL}$ and 
$V_{GR}$. Their alignment, schematically shown in Fig. 2b, results in a resonant current ridge defining the overlapping bases of the two triangles in Fig. 2a (green cross). Moving along this ridge corresponds to shifting both levels across the bias window while maintaining their reciprocal alignment. Moving away from the ridge results in a finite level detuning, $\epsilon_0$, between the ground-state levels.

An additional current ridge can be identified in Fig. 2a (orange cross) which we ascribe to a resonant tunneling process from the ground-state 
of $D_L$ to the first excited state of $D_R$ (Fig. 2c). From the distance between this ridge and the base line we extract an excitation energy of 7.4$\pm$0.4 meV. Due to interface effects \cite{Rahman2009}, this value is lower than expected for an As dopant in bulk silicon \cite{Aggarwal1965}, but still an order of magnitude larger than typical level spacings in semiconductor quantum dots. In addition, current between the ridges is mostly below the noise level, meaning that inelastic tunneling between the donor ground states is virtually absent. These remarkable features indicate that the donor ground states form a TLS very well isolated from its environment.

To further uncover the quantum properties of this TLS we shall now investigate the effect of microwave-driven charge oscillations between the dopant ground states.
The hybridization of their ground states, with tunnel coupling $\Delta$ results in a level anticrossing at $\epsilon_0 = 0$. The energy splitting of the lowest double-impurity levels is given by $\sqrt{\Delta^2 + \epsilon_0^2}$. In the limit of large detuning ($\vert \epsilon_0 \vert >> \Delta $) it reduces to $\epsilon_0$, and the dopant states are only marginally hybridized. The shared electron is strongly localized on either $D_L$ or $D_R$. The corresponding states, $\vert L \rangle$ and $\vert R \rangle$, define base states of a charge qubit. Coherent oscillations between $\vert L \rangle$ and $\vert R \rangle$ can be promoted with the aid of microwave photons generated by a gate voltage modulation in the gigahertz frequency range\cite{Oosterkamp1998}. These oscillations result in a measurable stationary current through the double donor.

\begin{figure}
\begin{center}
\includegraphics[width=\columnwidth]{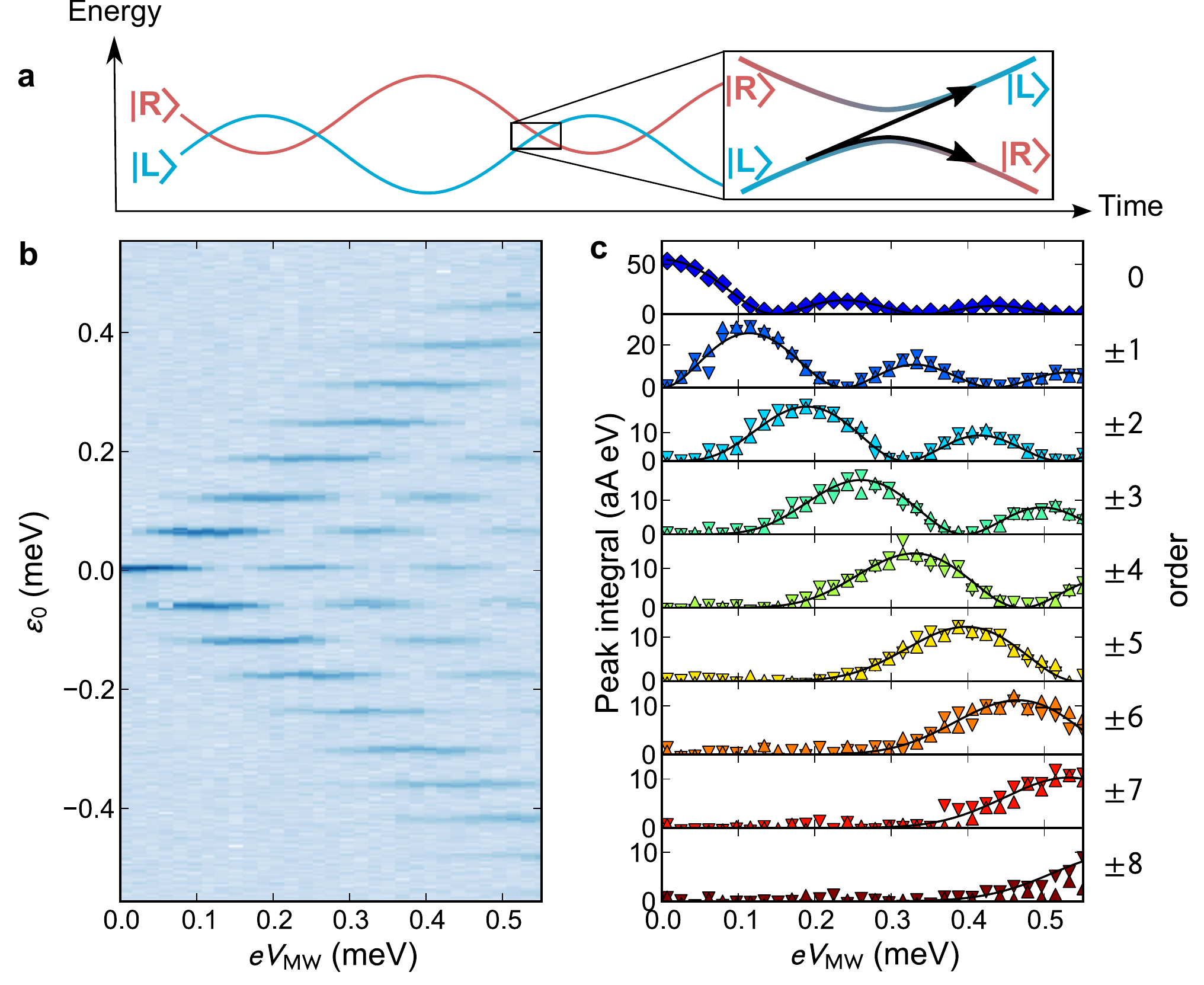}
\caption{\textbf{Landau-Zener-Stuckelberg interference pattern under microwave irradiation.} \textbf{a,} Time evolution, under microwave driving, of the two-level system (TLS) formed by the donor ground states.  $\vert L \rangle$ and $\vert R \rangle$ refer to the uncoupled donor ground states and $\Delta$ to their tunnel coupling amplitude. 
At $\epsilon_0 = 0$ the TLS eigenstates are even and odd combinations of  $\vert L \rangle$ and $\vert R \rangle$; for $\vert \epsilon_0 \vert >> \Delta$, they localize to  $\vert L \rangle$ and $\vert R \rangle$. At every passage through $\epsilon_0 = 0$ an "incoming" state $\vert L \rangle$ (or $\vert R \rangle$) splits into a superposition of ''outcoming''  $\vert L \rangle$ and $\vert R \rangle$ states. Staying in the same state requires a Landau-Zener transition across the gap $\Delta$. Multiple passages done within the dephasing time interfere with each other. 
\textbf{b,} Measured $I_{SD}(V_{MW},\epsilon_0)$ interference pattern for a microwave frequency $f_{MW}=15$ GHz and a source-drain bias $V_{SD}=5$ mV. $V_{MW}$, the amplitude of the voltage modulation between the two donors, is proportional to the externally applied microwave voltage (see below). The static detuning $\epsilon_0$ depends linearly on the position in the ($V_{GL}$,$V_{GR}$) plane. The scaling factor between gate-voltage and energy is adjusted to have the horizonal current ridges spaced by the photon energy. This factor agrees within 2\% with the one detetermined from the triangles in Fig. 2a.  
\textbf{c,} Current peak integrals extracted from \textbf{b}. Upward (downward) triangles refer to $n>0$ ($n<0$). The ensemble of 17 data sets is simultaneously fitted (black curves) using equation 1 integrated over peaks of order n, with three free parameters: $\Delta$, $\eta$, and the scaling factor for $V_{MW}$.
} 
\label{fig3}
\end{center}
\end{figure}

This effect is shown in Fig. 3b, where $I_{SD}$ is plotted as a function of microwave amplitude, $V_{MW}$, and level detuning $\epsilon_0$, for a microwave frequency $f_{MW}= 15$ GHz and $V_{SD} = 5$ mV. The ground state levels are positioned far away from Fermi levels of the leads.
The first noticeable feature is a set of equally spaced, horizontal current ridges positioned at  $\epsilon_0 = n h f_{MW}$, where $n=0$, $\pm 1$, $\pm 2$,... and $h$ is the Planck constant\cite{Stoof1996}. These ridges reflect tunneling currents assisted by the emission (for $\epsilon_0 > 0$) or the absorption (for $\epsilon_0 < 0$) of $\vert n \vert$ photons. We have observed well-defined ridges for up to $\vert n \vert = 10$ at $f_{MW}=$ 10GHz (see Fig. 4).

The second remarkable feature is a strong current modulation along each ridge. 
The resulting triangular-shaped pattern of current fringes is a quantum interference effect that can be explained as follows. The microwave field leads to an oscillatory time dependence of the level detuning, i.e. $\tilde{\epsilon_0}(t)= \epsilon_0 + e V_{MW} cos(2 \pi f_{MW} t)$, where $e$ is the electron charge. For $V_{MW} >  \epsilon_0$ (condition defining the triangular outline in Fig. 3b) the microwave field is large enough to drive the TLS through its level anti-crossing (at $\epsilon_0 = 0$) twice per microwave period $1/f_{MW}$. The relative time evolution of the driven energy levels is schematically depicted in Fig. 3a. 

After each tunneling event taking an electron from the source into $D_L$, the two-level system is in state $\vert L \rangle$.  At the first passage through $\epsilon_0 = 0$ the system can take two possible ''paths'', ending up either in state $\vert R \rangle$ or staying in the same state $\vert L \rangle$ (see inset to Fig. 3a). The latter possibility corresponds to a Landau-Zener transition across the energy gap of the TLS. Quantum mechanics allows for these two possible paths to interfere with each other at a following passage through the anticrossing. This effect, known as Landau-Zener-Stuckelberg  interference, is analogous to Mach-Zehnder interferometry for photons\cite{Oliver2005,Shevchenko2010}. Each passage through the avoided crossing driven by the microwave field is analogous to the passage of a photon through an optical beam splitter. During the time between two consecutive passages, the out-coming ''beams'', i.e. states  $\vert L \rangle$ and $\vert R \rangle$, acquire a quantum mechanical phase given by the time integral of the respective energies. Their interference will be constructive or destructive depending on whether their phase difference is an even or an odd multiple of $\pi$, respectively. This phase difference is an increasing function of the microwave amplitude, thereby accounting for the observed alternation of maxima and minima in the current along each ridge.

The overall Landau-Zener-Stuckelberg current pattern can be modeled by the expression (see Supplementary Material): 
\begin{widetext}
\begin{equation}
I_{SD}=\frac{e}{2}\sum_n\frac{\Delta^2 J_n^2}{h^2/4\pi^2 T_2+T_2(n h f_{MW}-\epsilon_0)^2+(1/\Gamma_R+1/2\Gamma_L)\Delta^2 J_n^2}
\end{equation}
\end{widetext}

where $J_n = J_n(eV_{MW}/hf_{MW})$ is the $n-$th order Bessel function of the first kind, $T_2$ is the dephasing time, $\Gamma_L$ ($\Gamma_R$) is the tunnel rate between dopant $D_L$ ($D_R$) and the source (drain) lead. 

At each ridge, $I_{SD}(\epsilon_0)$ is a Lorentzian function centered around $\epsilon_0 = n h f_{MW}$. Integrating this function for each $n$ value yields a set of 17 traces (Fig. 3c) which can be simultaneously fitted with only three free parameters: $\Delta$, $\eta \equiv T_2(1/\Gamma_R + 1/2 \Gamma_L)$, and the scaling factor between the externally applied microwave amplitude and $V_{MW}$ (this factor was already implicitly used to define the horizontal scale in Fig. 3b). We obtain $\Delta / h= 125 $ MHz  and $\eta =4.1$ ns$^2$. 

In order to determine $T_2$, these parameters where fed back into expression (1) and used to fit a high-resolution measurement of the resonant tunneling current $I_{SD}(\epsilon_0)$ in the absence of microwave excitation. We find $T_2 = 0.3$ ns, comparable to values reported for other types of charge qubits \cite{Nakamura2002,Hayashi2003,Petta2004}. 
We estimate that dephasing cannot be due to electrical noise through the filtered gate leads \cite{Barrett2003,Vorojtsov2005}. Instead it could arise from the escape tunneling time\cite{Hayashi2003}  ${\Gamma_R}^{-1}$  or from the noise coming from switching 
offset charges \cite{Nakamura2002,Yurkevich2010}.

\begin{figure}
\begin{center}
\includegraphics[width=0.8\columnwidth]{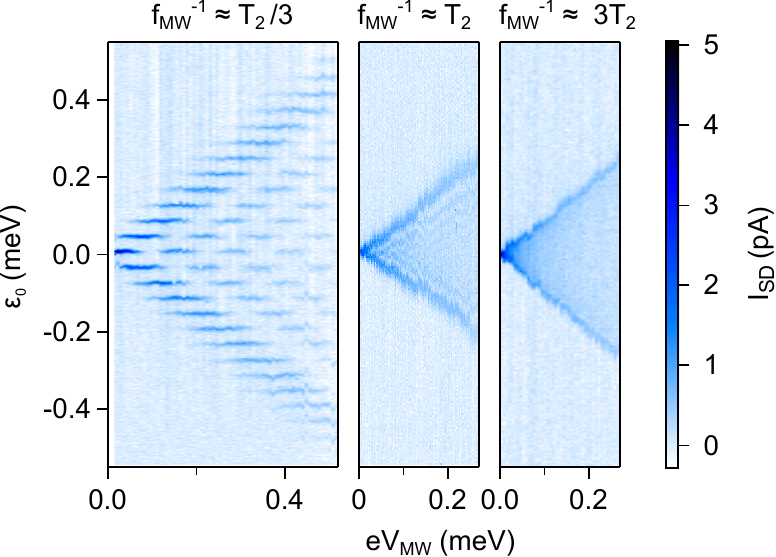}
\caption{\textbf{Transition from coherent to incoherent driving of the double-donor TLS.} Current maps for microwave driving frequencies from well above to well below the dephasing rate ${T_2}^{-1}$. \textbf{a}, For $f_{MW} = 10 $GHz$ \approx 3/T_2$ we observe sharply defined interference fringes that can be seen as the superposition of two interference patterns: one consisting of diagonal and anti-diagonal stripes, issued from the interference between two consecutive passages through $\epsilon_0 = 0$, and one consisting of horizontal stripes, issued from the constructive interference between consecutive pairs of passages through $\epsilon_0 = 0$. In the latter case, the time delay between consecutive pairs equals ${f_{MW}^{-1}}$ and constructive interference is equivalent to having a static detuning equal to an integer number of photon quanta. \textbf{b}, For $f_{MW} = 3 $GHz $ \approx T_2^{-1}$, coherence is preserved on the time scale of just one microwave period. As a result, only the interference stripes due two consecutive passages through $\epsilon_0 = 0$ remain visible. \textbf{c}, For $f_{MW} = 1 $GHz $ \approx (3T_2)^{-1}$, coherence is lost also between consecutive passages through $\epsilon_0 = 0$ leading to a structureless current map.
}
\label{fig4}
\end{center}
\end{figure}

Our evaluation of $T_2$ is confirmed by the frequency dependence of the Landau-Zener-Stuckelberg pattern\cite{Shevchenko2010}. Upon increasing the microwave period $f_{MW}^{-1}$ from well below to well above $T_2$, three distinct regimes can be identified, as shown by the data sets in Fig. 4. For $f_{MW}^{-1} \approx T_2/3$ (left panel), multiple passages through the level anticrossing occur within the coherence time, leading to clearly defined and well-separated interference fringes as the ones shown in fig3.  For $f_{MW}^{-1} \approx T_2$ (middle panel), only two consecutive passages are allowed within $T_2$ and successive microwave periods are uncorrelated. As a result, the interference fringes blur and the nodes of the Bessel functions merge into lines departing from the zero detuning point. Finally, for $f_{MW}^{-1} \approx 3 T_2$ (right panel), even consecutive passages become uncorrelated leading to an average current with no structure. These measurements clearly show the transition from quantum coherent to incoherent driving of the double-donor charge qubit.

Because of the large energy separation between the donor's ground and excited states, we argue that driving signals of much higher-frequency (in principle up to $f_{MW} = 1$ THz, i.e. hundreds of times larger than the decoherence rate) could be used without inducing unwanted excitations. Furthermore, recent progress in the positioning of individual donors with nanometer precision\cite{Fuechsle2012} will give access to fine tuning of tunnel coupling between the two donors. This high-frequency driving combined with a large tunnel coupling should provide the best configuration for the development of donor-based qubits.

\flushleft{\tiny\bf Acknowledgements}\\
\flushleft{\scriptsize We thank Y. Nazarov, M. Houzet, J. Meyer and J.-P. Brison for useful discussions. This work was supported by the Agence Nationale de la Recherche (ANR) through the ACCESS project and by the European Commission through the FP7 FET-proactive NanoICT and the European Starting Grant programmes. 
 }\\
\flushleft{\tiny\bf Author contributions } \\
\flushleft{\scriptsize E. D.-F. set up the equipment for microwave experiments at cryogenic temperatures. E. D.-F., B. R., and B.V. carried out the experiments and analyzed the data under the supervision of X.J., M.S., and S.D.F.  R.W. and M.V. supervised the device fabrication process. All authors co-wrote the manuscript.  }\\

\newpage

\end{document}